\documentclass[sigconf,authorversion]{acmart}

\usepackage{algorithm}
\usepackage{algorithmic}

\AtBeginDocument{%
  \providecommand\BibTeX{{%
    \normalfont B\kern-0.5em{\scshape i\kern-0.25em b}\kern-0.8em\TeX}}}

\copyrightyear{2020}
\acmYear{2020}
\setcopyright{acmlicensed}\acmConference[ASE '20]{35th IEEE/ACM International Conference on Automated Software Engineering}{September 21--25, 2020}{Virtual Event, Australia}
\acmBooktitle{35th IEEE/ACM International Conference on Automated Software Engineering (ASE '20), September 21--25, 2020, Virtual Event, Australia}
\acmDOI{10.1145/3324884.3416549}
\newcommand{\causality}[3]{
    \begin{center}
        \begin{tabular}{|r|p{6cm}|}
            \hline
            \textbf{Sentence} & #1 \\ \hline
            \textbf{Causality} & #2 $\Longrightarrow$ #3 \\ \hline
        \end{tabular}
    \end{center}
}

\begin{document}

\title{Automatic Extraction of Cause-Effect-Relations from Requirements Artifacts}


\author{Julian Frattini}
\email{julian.frattini@bth.se}
\orcid{0000-0003-3995-6125}
\affiliation{
  \institution{Blekinge Institute of Technology}
  \streetaddress{Valhallavägen 1}
  \city{Karlskrona}
  \state{Sweden}
  \postcode{37140}
}

\author{Maximilian Junker}
\email{maximilian.junker@qualicen.de}
\affiliation{
  \institution{Qualicen GmbH}
  \streetaddress{Lichtenbergstr. 8}
  \city{Munich}
  \state{Germany}
  \postcode{85748}
}

\author{Michael Unterkalmsteiner}
\email{michael.unterkalmsteiner@bth.se}
\affiliation{
  \institution{Blekinge Institute of Technology}
  \streetaddress{Valhallavägen 1}
  \city{Karlskrona}
  \state{Sweden}
  \postcode{37140}
}

\author{Daniel Mendez}
\email{daniel.mendez@bth.se}
\orcid{0000-0003-0619-6027}
\affiliation{
  \institution{Blekinge Institute of Technology and fortiss GmbH}
  \streetaddress{Valhallavägen 1}
  \city{Karlskrona}
  \state{Sweden}
  \postcode{37140}
}

\renewcommand{\shortauthors}{Frattini, et al.}

\begin{abstract}
\textit{Background:} The detection and extraction of causality from natural language sentences have shown great potential in various fields of application. The field of requirements engineering is eligible for multiple reasons: (1) requirements artifacts are primarily written in natural language, (2) causal sentences convey essential context about the subject of requirements, and (3) extracted and formalized causality relations are usable for a (semi-)automatic translation into further artifacts, such as test cases. \newline
\textit{Objective:} We aim at understanding the value of interactive causality extraction based on syntactic criteria for the context of requirements engineering. \newline
\textit{Method:} We developed a prototype of a system for automatic causality extraction and evaluate it by applying it to a set of publicly available requirements artifacts, determining whether the automatic extraction reduces the manual effort of requirements formalization. \newline
\textit{Result:} During the evaluation we analyzed 4457 natural language sentences from 18 requirements documents, 558 of which were causal (12.52\%). The best evaluation of a requirements document provided an automatic extraction of 48.57\% cause-effect graphs on average, which demonstrates the feasibility of the approach. \newline
\textit{Limitation:} The feasibility of the approach has been proven in theory but lacks exploration of being scaled up for practical use. Evaluating the applicability of the automatic causality extraction for a requirements engineer is left for future research. \newline
\textit{Conclusion:} A syntactic approach for causality extraction is viable for the context of requirements engineering and can aid a pipeline towards an automatic generation of further artifacts from requirements artifacts.
\end{abstract}





\maketitle

\section{Introduction}
\label{section:intro}
The detection of semantic relations in natural language text has proven to be an aspect of the field of information extraction which has great potential for many areas of application as outlined by the SemEval 2010 task set \cite{hendrickx2010semeval}\cite{verhagen2010semeval}\cite{recasens2010semeval}. \newline
The automatic extraction of causal relations from large corpora of natural language text has proven useful to many fields of research. Some examples for these fields of research are healthcare, where causal relations between symptoms and diseases yield insights for diagnostics \cite{khoo2000extracting}, and economy, where financial relations can be derived from analysing stock reports \cite{chan2005extracting}. \newline
These domain-specific examples limit the scope of an automatic phrase extraction method to causal relations but they in turn increase their precision, as specific semantic structures and lexical cue phrases reoccur in domain-specific texts. Automatic extraction of causal relations is especially applicable in domains where causal relations are explicitly stated. \newline
One field of research that is often overlooked in this context, but shares the qualifying attributes to excel in automated recognition and extraction of causal relations, is requirements engineering. Artifacts in requirements engineering are predominantly written in natural language \cite{wagner2019status} and describe conditions of a system in sentences conveying a causality relation. This can be seen in the example "If registration is not successful an audible and visual indication shall be provided." \cite{ferrari2017pure}, where the relation between the unsuccessful registration and the audible and visual indication is expressed in a causal sentence. \newline
Furthermore, the implementation of a system based on requirements engineering artifacts needs to be validated by determining whether the defined requirements are fulfilled. Tests of various granularity are usually a formalized version of specific requirements. Manually transforming natural language requirements into formalized test cases can be a tedious and error-prone task, as natural language often lacks specificity and contains ambiguous phrasing, which may be interpreted in more than one way. \newline
Approaches to minimize ambiguity in natural language requirements exist in the form of controlled natural language \cite{fuchs1995controlled}\cite{mavin2009ears}, which can be easily reused for further formalization \cite{selway2015formalising} or generating executable scenarios from requirements \cite{gordon2009generating} among others. Though controlled natural language poses clear advantages for formalization techniques, it is still underrepresented in requirements engineering in practice. For this reason we do not assume any form of control on the natural language requirements. \newline
An automatic recognition and extraction of causal relations in natural language requirements texts can mitigate the problem of ambiguity by reducing the manual work necessary for transforming causal relations conveyed by sentences into formalized notation. \newline
Above that recent studies have shown that an automatic test generation from cause-effect graphs are effective and viable \cite{srivastava2009cause}\cite{son2014test}. Our proposed system fills the gap between natural language requirements artifacts and cause-effect graphs: the automatic extraction of causal relations may be used in a pipeline to automatically generate test cases from natural language requirements artifacts. \newline
Our contribution is threefold: first, we introduce a causality pattern structure that is based on the syntactic rather than the semantic structure of a sentence. This structure is specified incrementally to prevent over-fitting. Second, we present an interactive, online machine learning framework that tailors the causality patterns to its input, therefore adapting to any given context. Third, we evaluate the framework by analysing a set of publicly available requirements artifacts and assess the system's capability to automatically extract cause-effect graphs from causal sentences. \newline
Following Roel Wieringa's design science approach \cite{wieringa2014design} our work relies on the problem investigation observed in practice in the industry, where requirements engineering experts at Qualicen GmbH identified a need for an automatic extraction of causal relations from natural language requirements. Our contribution covers the treatment design and provides a preliminary evaluation of the prototypical implementation. \newline
The rest of this paper is structured as follows: Section~\ref{section:related} covers related work in the field of automatic extraction of causal relations. Section~\ref{section:concepts} introduces the used concepts and Section~\ref{section:impl} presents our approach in detail, which is evaluated in Section~\ref{section:eval}. A conclusion is drawn in Section~\ref{section:conclusion} alongside considerations for future work.

\section{Related Work}
\label{section:related}
% causality extraction
Automatic extraction of causal relations from natural language texts is typically split into two categories \cite{asghar2016automatic}: non-statistical techniques that rely on pattern matching and statistical techniques that utilize machine learning for classification of sentences. \newline
Early approaches attempted an automatic causality detection by matching sentences to manually defined lexico-syntactic patterns. A highly cited, early work by Girju et al. \cite{girju2002text} investigated a specific pattern <NP\textsubscript{1} VP NP\textsubscript{2}>, where NP represented a noun phrase and VP a verb phrase. A set of semantic constraints on the three pattern elements filtered the matching sentences and allowed for a causality detection via a specific, semantic ranking. Other pattern-based approaches include \cite{chan2005extracting}, which utilizes a hierarchy of templates to increase the granularity of patterns. \newline
Another popular approach consists of feature-based classification methods, where causality is detected based on syntactic, semantic or lexical features. A recent approach by Ayyanar et al. \cite{ayyanar2019causal} based the classification on grammar tags of specific elements of a sentence as well as the distance between the causally related nouns. Other approaches on classification via features and decision trees are \cite{blanco2008causal}, where syntactic patterns were manually generated and then used to train a decision tree for the detection of causality, and \cite{girju2003automatic}, a refinement of \cite{girju2002text} by adding a decision tree for causality detection. \newline
Rink et al. achieved the best result for the SemEval 2010 task 8 challenge of detecting causality \cite{asghar2016automatic}. Their proposed framework combines graphical patterns consisting of syntactic, semantic and lexical constraints with a binary classifier to identify causally related events in sentences \cite{rink2010learning}. \newline
% causality extraction in RE
Prominent domains for application of causality extraction are question answering \cite{girju2003automatic}, the medical domain \cite{khoo2000extracting}, and economics \cite{chan2005extracting}. \newline
The technique can be utilized in the context of requirements and software engineering to reuse extracted cause-effect graphs for other downstream artifacts. One example of this is to integrate an automatic extraction algorithm into a pipeline for automatic test case generation: As test cases are the validation of a built system and this validation is based on the context described in the requirements artifacts \cite{utting2012taxonomy}, an automatic pipeline for formalizing natural language requirements artifacts into test cases provides viable support for the software engineering process \cite{paradkar1997ceg}. \newline
Research on the steps of automatically transforming cause-effect graphs into decision tables \cite{srivastava2009cause}\cite{son2014test} exists as well as transforming decision tables into test suites \cite{sharma2010automatic}. \newline
Other possible fields of application include traceability link recovery, where the extracted semantic, causal relation could add to the processing of natural language requirements \cite{lapena2018exploring}. The extracted cause-effect graph provides additional semantic information on relations in the requirements and might add to the precision of link recovery. \newline
But most of the existing approaches on causality extraction are not eligible to the context of requirements engineering for three major reasons:
\begin{enumerate}
    \item Causal relations described in requirements artifacts are rarely connected in an obvious semantic manner
    \item Semantic causality extraction is based on large corpora of natural language text, which is often not available in the requirements engineering phase
    \item The extracted causal relation is often limited to a word pair, omitting information of the causal sentence vital for reusing the extracted cause-effect graph
\end{enumerate}
Tailoring causality extraction for requirements engineering is explored in a more recent approach by Fischbach et al. \cite{fischbach2019automated}, where test cases are automatically generated from acceptance criteria. The usage of predefined, manual patterns based on the dependency structure of sentences specifies their approach on the artifact type of user stories, which is avoided in our approach to ensure a greater generalizability. \newline
Our approach is eligible to extract cause-effect graphs from causal sentences in requirements engineering by dealing with the aforementioned flaws of existing approaches as follows:
\begin{enumerate}
    \item Causal relations are detected via syntactic and lexical attributes
    \item The online learning approach continuously improves the causality detection without the need for a large training corpus beforehand
    \item Specific phrase extraction methods are capable of extracting an arbitrary portion of a sentence rather than only a single word.
\end{enumerate}
This approach can be applied to requirements engineering as well as any other domain.

\section{Concepts}
\label{section:concepts}
This section introduces and formalizes all concepts used in the context of the implementation.

% Foundation of concepts: sentences and causality
\subsection{Foundation} 
\label{section:concepts:foundation}

\paragraph{Sentences}
We denote sentences as $s \in S$, where S is the set of all sentences, called the corpus, and a sentence s can be causal or non-causal.

\paragraph{Causality}
Causal sentences contain a causality $c \in C$. We introduce the following predicate to express the causality of a sentence s:
\begin{center}
    \begin{tabular}{|l|p{5cm}|} \hline
        \textbf{Predicate} & \textbf{Explanation} \\ \hline
        causality(s, c) & $c \in C$ is the causality of sentence $s \in S$ \\ \hline
    \end{tabular}
\end{center}
A causal sentence and its conveyed causality is illustrated in the following example:
\causality{The application is terminated when the x-button is pressed.}{the x-button is pressed}{The application is terminated}
The relation of causality is reduced to its most simple form, consisting of a cause- and an effect-phrase. Extracting these two phrases from the sentence is subject to the causality extraction.

% Formalization of concepts: structure, structural equivalence, internal representation and cause-effect-graphs
\subsection{Formalization} 
\label{section:concepts:formaliztion}

\paragraph{Structure}
A sentence written in natural language can be processed by natural language processing (NLP) tools in order to derive syntactic and semantic information about the sentence. A sentence can be structured in two ways: by constituency or dependency. A constituency parser groups adjacent words into grammatical units and builds a tree structure, where the root node represents the full sentence. A dependency parser also constructs a tree structure, but in this case the inner nodes of the tree are not chunks grouping words or other chunks into units, but rather the words itself. A word $w_1$ is the parent node of a word $w_2$ if $w_2$ is semantically dependent of $w_1$. The root node of the semantic structure is the root dependency, which is usually the sentence's predicate. \newline
Elemental to the identification of causal sentences in our approach is the syntactic structure, which will be denoted as $t \in T$ and applicable to the following predicates:
\begin{center}
    \begin{tabular}{|l|p{5cm}|} \hline
        structureOf(s, t) & $t \in T$ is the structure of sentence $s \in S$ \\ \hline
    \end{tabular}
\end{center}

\begin{figure*}
    \centering
    \includegraphics[width=\textwidth]{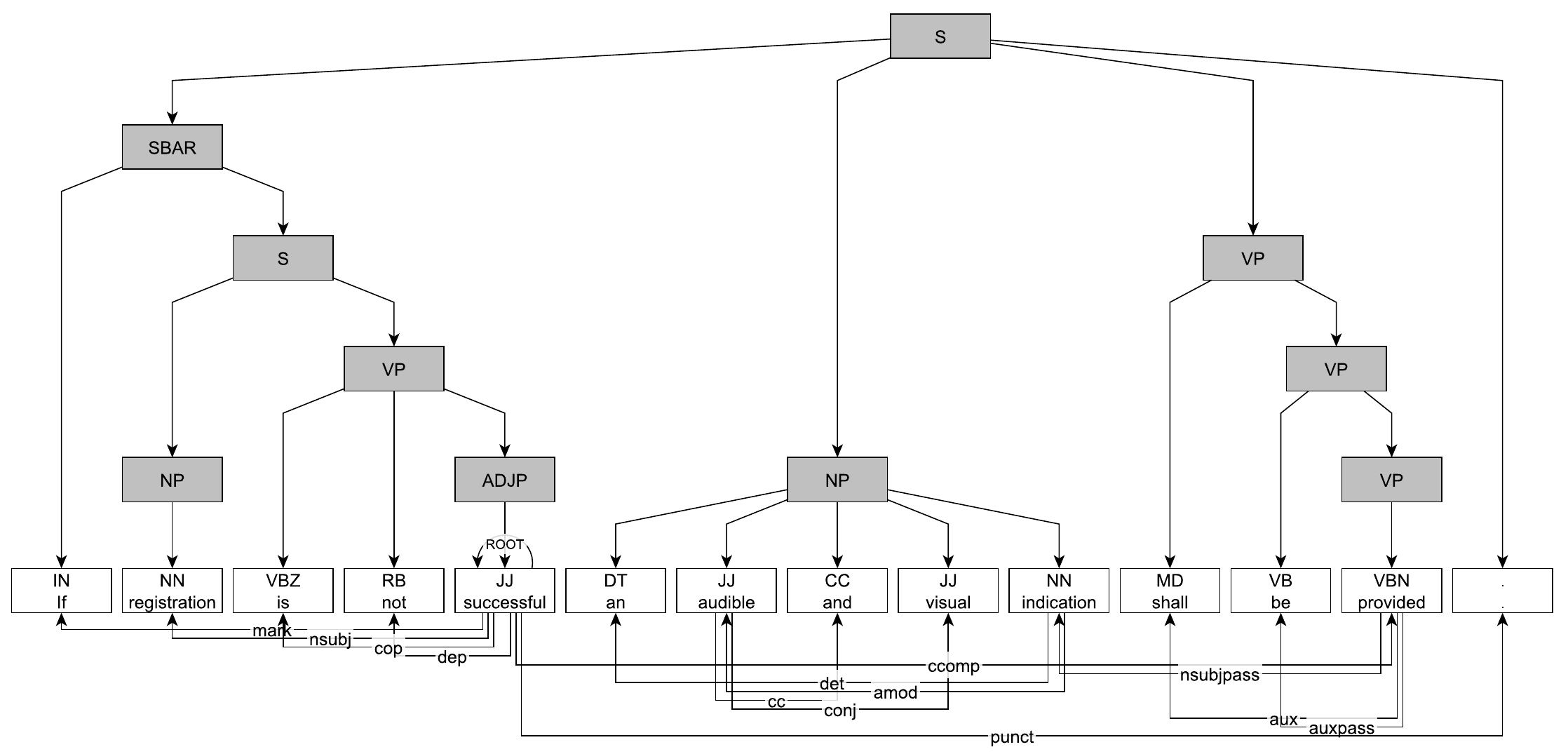}
    \caption{Internal representation of a natural language sentence formalized via constituency and dependency parsing}
    \label{fig:internalRepresentation}
\end{figure*}

\paragraph{Internal Representation}
A sentence $s \in S$ is formalized by NLP tools into an internal representation. This internal representation of sentences consists of the following aspects:
\begin{itemize}
    \item Words: each word contained in the sentence in order of appearance with a corresponding part-of-speech-tag indicating the words role within the sentence
    \item Constituency structure: adjacent words clustered together to chunks of words which represent a specific grammatical unit
    \item Dependency structure: semantic relations of words to each other
\end{itemize}
An example for a sentence formalized into the internal representation is given in Figure~\ref{fig:internalRepresentation}. The tree-structure composed of grey nodes represents the constituency structure, where each node contains its constituent tag. The list of white nodes represents the words of the sentence annotated with their respective part-of-speech-tag. The lines connecting the word nodes represent the dependency structure, where each label on the connection represents the dependency relation type.

\begin{figure*}
    \centering
    \begin{minipage}{0.3\textwidth}
        \centering
        \includegraphics[width=\textwidth]{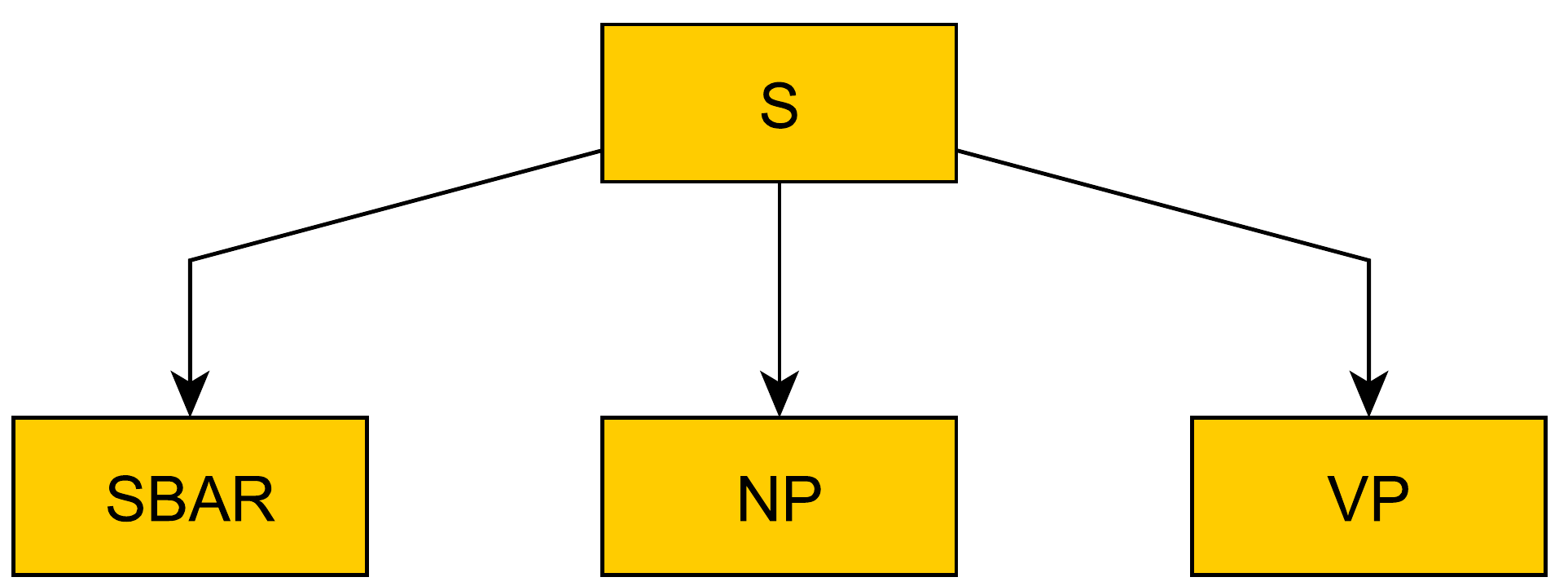}
        \caption{Syntactic sentence structure $t_1$ of sentence $s_1$}
        \label{fig:structure1}
    \end{minipage}\hfill
    \begin{minipage}{0.3\textwidth}
        \centering
        \includegraphics[width=\textwidth]{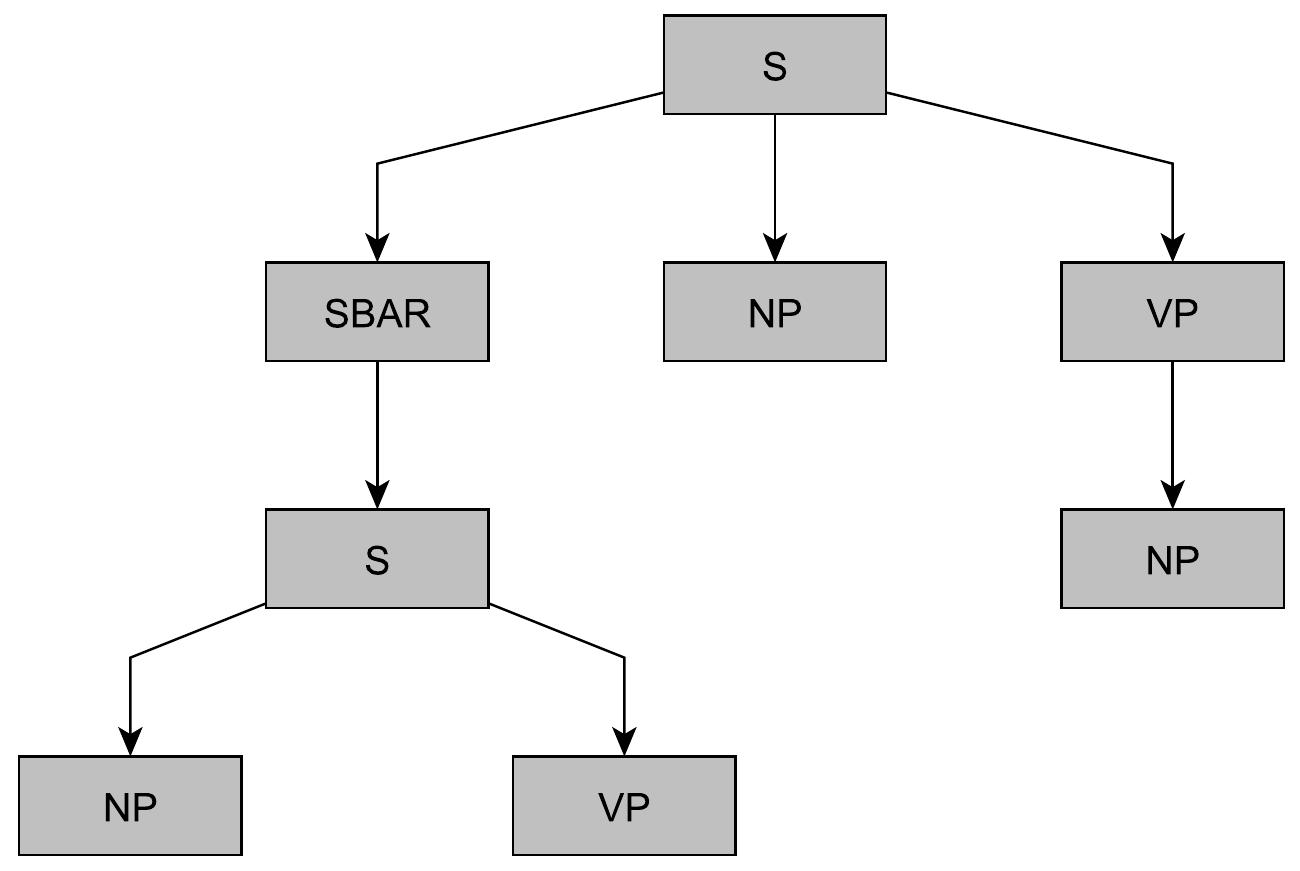}
        \caption{Syntactic sentence structure $t_2$ of sentence $s_2$}
        \label{fig:structure2}
    \end{minipage}\hfill
    \begin{minipage}{0.3\textwidth}
        \centering
        \includegraphics[width=\textwidth]{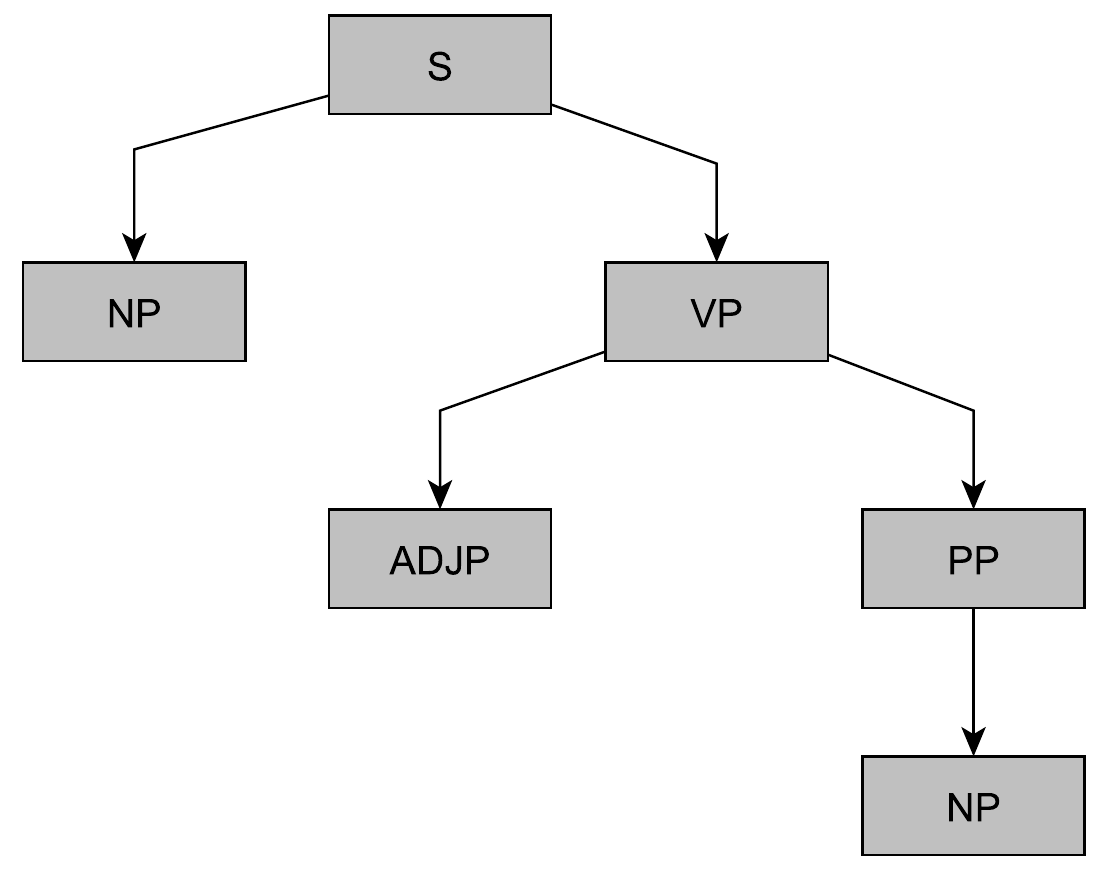}
        \caption{Syntactic sentence structure $t_3$ of sentence $s_3$}
        \label{fig:structure3}
    \end{minipage}
\end{figure*}

\paragraph{Structural Equivalence}
The premise of our approach is that similarly structured causal sentences also have their cause- and effect-phrases located at similar positions within the sentence. This premise will be explained in detail in Section~\ref{section:impl:maintenance}. Two tree-like structures are equivalent if both structures contain the same nodes in the same order. Furthermore, a structure $t_1 \in T$ is the subtree of a structure $t_2 \in T$ when $t_1$ contains all nodes in the same order as $t_2$ and potentially more, but not necessarily the other way around. We introduce the following predicate:
\begin{center}
    \begin{tabular}{|l|l|p{5cm}|} \hline
        subtree($t_1$, $t_2$) & structure $t_1 \in T$ is a subtree of structure $t_2 \in T$ \\ \hline
    \end{tabular}
\end{center}
For example, the structure depicted in Figure~\ref{fig:structure1} is a subtree of the structure in Figure~\ref{fig:structure2}, but not a subtree of structure in Figure~\ref{fig:structure3}. Hence: $subtree(t_1, t_2)$ but not $subtree(t_1, t_3)$, because $t_3$ lacks a node with the tag SBAR as the first child of the root node.

\paragraph{Cause-Effect Graphs}
Cause-effect graphs are used to represent the causality, which a sentence conveys, in a formalized, graphical notation \cite{elmendorf1973cause}. In our approach we limit the extent of cause-effect graphs to their most simple form, consisting of two nodes, one representing the cause and one the effect of the causality relation. Cause-effect graphs will be denoted as $g \in G$ and are applicable to the following predicate:
\begin{center}
    \begin{tabular}{|l|p{5cm}|} \hline
        conveys(g, c) & CEG $g \in G$ conveys causality $c \in C$ \\ \hline
    \end{tabular}
\end{center}
For every causal sentence there exists a cause-effect graph that conveys its causality:
$$ \forall s \in S, \exists c \in C : causality(s, c) \rightarrow \exists g \in G : conveys(g,c) $$
Note that the cause-effect graph is not necessarily equivalent to the causality. Ensuring that a cause-effect graph extracted from a sentence actually represents the causality of that sentence is the center of the challenge of correct causality extraction.

\subsection{Phrase Extraction Methods}
\label{section:concepts:extractionAlgorithms}
Other attempts at causality extraction covered the part of phrase extraction by selecting certain semantic elements from the sentence, for example taking the two noun phrases NP\textsubscript{1} and NP\textsubscript{2} of the sentence. Introducing phrase extraction methods is necessary for our approach for two reasons: first, to provide independence of semantic information, the extraction of phrases must be based on syntactic or lexical information. Second, the extent of the phrase, which represents a part of the causality, shall be arbitrary to ensure that the causality patterns meet the needs for the requirements engineering process. An example from the prominent SemEval 2010 Task 8 set of annotated sentences is the following sentence \cite{hendrickx2010semeval}:
\causality{The current view is that the chronic inflammation in the distal part of the stomach caused by Helicobacter pylori infection results in an increased acid production from the non-infected upper corpus region of the stomach.}{infection}{inflammation}
The causality relation annotated in the SemEval 2010 set of sentences is semantically correct, but causality detection algorithms trimmed to identifying causal relations of this type are restricted to ignore a lot of data in comparable sentences. Without claiming to define the borders of causality in a linguistic or philosophical scope, it is feasible to expect a more encompassing phrase from a causality phrase extraction method, for example as follows:
\causality{The current view is that the chronic inflammation in the distal part of the stomach caused by Helicobacter pylori infection results in an increased acid production from the non-infected upper corpus region of the stomach.}{Helicobacter pylori infection}{chronic inflammation in the distal part of the stomach}
The context of causality extraction in requirements engineering is to extract phrases from a natural language requirements artifact that can be reused to generate test cases, possibly even automatically. The information value of an extracted causality is highest when the extraction is not limited to the two events in a causal relation represented by a single word, but includes all relevant information like participating actors and conditions. This becomes evident in the following example, where a sentence from the PURE dataset of requirements artifacts \cite{ferrari2017pure} is annotated by a causality, which only takes into account the connected events:
\causality{If registration is not successful an audible and visual indication shall be provided.}{registration}{indication}
The causal relation of the registration and the indication is evident, but vital information is lost in the extraction process. This vital information is necessary to more precisely define the initial situation and the expected result of described context. In contrast, the following cause- and effect-phrase would provide the necessary information for later processing, like a test case generation:
\causality{If registration is not successful an audible and visual indication shall be provided.}{registration is not successful}{an audible and visual indication shall be provided}
An phrase extraction method $e \in E$ is a process that, when applied to a sentence $s \in S$, returns a specific phrase from this sentence. The extracted phrase is a substring of the sentence. An phrase extraction method is defined as follows: for a sentence s with causality c, the application of the phrase extraction method on s will yield a cause-effect graph g. The following predicate is introduced for phrase extraction methods $e \in E$:
\begin{center}
    \begin{tabular}{|l|p{5cm}|} \hline
        extracts(e, s, g) & Phrase extraction method $e \in E$ extracts cause-effect graph $g \in G$ from sentence $s \in S$ \\ \hline
    \end{tabular}
\end{center}
A phrase extraction method is generated as outlined in Algorithm~\ref{alg:extraction}. In the simplified case of cause-effect graphs for this approach the cause-effect graph g only contains 2 nodes n, one for the cause and one for the effect. 

\begin{algorithm}
    \caption{Generate phrase extraction method $e \in E$ for causal sentence $s \in S$ with causality $c \in C$}
    \label{alg:extraction}
    \begin{algorithmic} 
        \REQUIRE $s \in S$, cause-effect graph $g \in G$ where causality(s, c)
        \FOR{node $n \in g $} 
            \STATE find phrase of n in s
            \STATE find nodes in the syntactic structure of s, which parent the phrase of n
            \STATE create selector of these nodes and add selector to e
        \ENDFOR
        \RETURN $e$, where extracts(e, s, g) holds
    \end{algorithmic}
\end{algorithm}

The resulting phrase extraction method works by selecting specific nodes within a sentence and combining all word nodes, that are direct or indirect children of the selected node, together. Multi-word expressions are handled by selecting parenting nodes of the sentence structure, which cover the full expression. \newline
In case of the sentence described above and shown in Figure~\ref{fig:internalRepresentation}, the phrase extraction method extracting the cause-phrase would select the node S under the node SBAR, which transitively parents all word nodes of which the cause-phrase consists. The phrase extraction method extracting the effect-phrase would select the node NP and VP, which are direct children of the root node, and combine their word nodes to the effect-phrase.

\subsection{Causality Patterns}
\label{section:concepts:patterns}
A causality pattern combines the recognition of a causal sentence with extraction of a cause- and effect-phrase from this unknown sentence. This connection is based on the grammatical equivalence to a previously known and processed causal sentence example. A causality pattern consists of three elements:
\begin{itemize}
    \item Signature: a subtree of a syntactic sentence structure, which is the indicator for compliance to a sentence
    \item Phrase extraction method: an algorithm, which extracts the cause- and effect-phrase when applied to a sentence
    \item Accepted sentences: a list of all causal sentences, that are compliant to the pattern and applicable by its phrase extraction method
\end{itemize}
%In the case of simplified cause-effect graphs as described in Section~\ref{section:concepts:formaliztion} a causality pattern must be associated with exactly two phrase extraction methods: one for extracting the cause-phrase, one for extracting the effect-phrase. 
The signature associated with a causality pattern is similar to the syntactic structure $t \in T$ of a sentence, as it represents a sub-tree of a syntactic structure. \newline
The following predicates are introduced:
\begin{center}
    \begin{tabular}{|l|p{5.5cm}|} \hline
        signatureOf(p, t) & $t \in T$ is the signature of the pattern $p \in P$ \\ \hline
        compliant(p, s) & sentence $s \in S$ is compliant to pattern $p \in P$ \\ \hline
        extractionOf(p, e) & $e \in E$ is the phrase extraction method of pattern $p \in P$ \\ \hline
        applicable(p, s) & sentence $s \in S$ is applicable by the pattern $p \in P$ \\ \hline
        accepted(p, s) & sentence $s \in S$ is accepted by the pattern $p \in P$ \\ \hline
    \end{tabular}
\end{center}

\paragraph{Generation}
A causality pattern can be generated from a causal sentence and a manually created cause-effect graph. The signature is extracted from the sentence's syntactic structure to represent its grammatical structure, and the phrase extraction methods are generated from the elements of the cause-effect graph, which are located within the sentence. The generation and maintenance is further explained in Section~\ref{section:impl:maintenance}. The acceptance of a sentence by a pattern is determined by evaluating compliance and applicability, which are explained next. 

\paragraph{Compliance} Compliance between a sentence and a pattern determines, whether the sentence belongs to the grammatical equivalence class which the pattern represents. A sentence is compliant to a pattern if and only if the signature of the pattern is a subset of the sentence structure. This is formally expressed as:
\begin{multline*}
    \forall s \in S, \forall p \in P, \exists t_1, t_2 \in T : signatureOf(p, t_1) \land \\ 
    structureOf(s, t_2) \land subtree(t_1, t_2) \rightarrow compliant(p, s)
\end{multline*}

\begin{figure}
    \centering
    \includegraphics[width=0.25\textwidth]{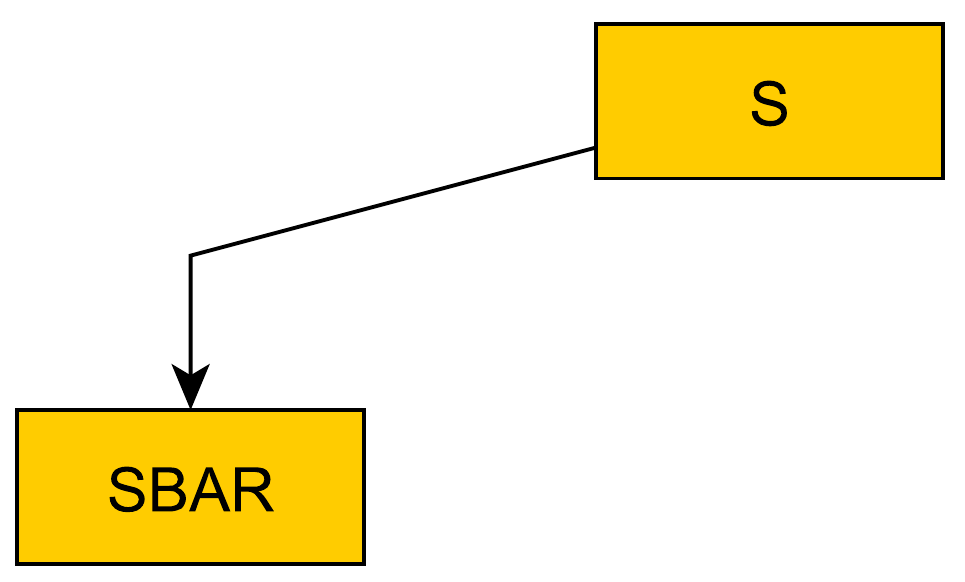}
    \caption{Signature $t_{p2}$ of a hypothetical pattern $p_2$}
    \label{fig:signature1}
\end{figure}

Our approach explores an incremental signature: the pattern's signature consists of as many nodes necessary, such that the following two conditions hold:
\begin{itemize}
    \item The pattern's signature is a subtree to the structure of all accepted sentences
    \item The pattern's signature is not a subtree to the structure of all non-causal sentences and causal sentences of different patterns.
\end{itemize}
This can be visualized when reviewing existing examples: assume sentence $s_2$ with structure $t_2$ visualized in Figure~\ref{fig:structure2} is causal and sentence $s_3$ with structure $t_3$ visualized in Figure~\ref{fig:structure3} is not causal. Then a pattern $p_2$ would minimally require a signature $t_{p2}$ as visualized in Figure~\ref{fig:signature1} to differentiate the two sentences. Since $t_{p2}$ is a subtree of $t_2$, sentence $s_2$ is compliant to pattern $p_2$. Sentence $s_3$ is not compliant to pattern $p_2$, because $t_{p2}$ is not a subtree of $t_3$. Note that the pattern $p_2$ would also differentiate between $s_2$ and $s_3$ if it had a the signature $t_1$ shown in Figure~\ref{fig:structure1}, but this signature would not be minimal because it contains more nodes than necessary. The conditions ensure that only sentences of similar syntactic sentence structure are compliant to a pattern and therefore eligible to be accepted. 

\paragraph{Applicability}
Applicability of a sentence by a pattern determines, whether the phrase extraction methods of the pattern extract the desired cause-effect graph. A sentence is applicable by a pattern if the phrase extraction method associated with the pattern extracts the cause- and effect-expressions from the sentence that convey its causality relation. This is formally:
\begin{multline*}
    \forall p \in P, \forall s \in S, \exists e \in E, \exists c \in C, \exists g \in G : causalityOf(s, c) \land \\ 
    conveys(g, c) \land extractionOf(p, e) \land extracts(e, s, g) \\ 
    \rightarrow applicable(p, s)
\end{multline*}

\paragraph{Acceptance}
A sentence is accepted by a pattern if the sentence's structure is compliant to the pattern's signature and the sentence is applicable by the patterns phrase extraction method.
$$\forall s \in S, \forall p \in P : compliant(p, s) \land applicable(p, s) \rightarrow accepted(p, s)$$
\section{Implementation}
\label{section:impl}

This Section describes the implementation of concepts and the functionality of the pattern-based machine learning algorithm, which will then be evaluated in the following chapter. We invite fellow researchers to inspect and use our implementation \cite{cerec2020} for further studies.

\subsection{System Architecture}
The \textbf{c}ause-\textbf{e}ffect \textbf{rec}ognition system (cerec) provides an interface which allows for two operations: \textit{training}, where the system is provided with a sentence and a desired cause-effect graph, which conveys the sentence's causality. Here the system is induced to create a new causality pattern for the sentence's syntax, if it does not already exist. When using the operation \textit{testing} the system is provided only with a sentence. If the sentence is causal and its syntactic structure is already covered by a causality pattern, the system will attempt to generate a cause-effect graph from the sentence using the phrase extraction methods associated with the pattern. If a sentence is not correctly processed -- meaning that a causal sentence is either not recognized as causal or a non-causal wrongfully recognized as causal -- providing a manual correction will result in a training operation to generate a new or adapted pattern. \newline
The \textit{cause-effect recognition} is the core of the cerec system and tasked with the maintenance of pattern correctness, which is in detail explained in Section~\ref{section:impl:maintenance}. This system contains three major elements:
\begin{itemize}
    \item \textit{NLP} unit: used to formalize natural language sentences into the internal representation mentioned in Section~\ref{section:concepts:formaliztion}
    \item \textit{CommandGenerator}: used to generate phrase extraction methods mentioned in Section~\ref{section:concepts:extractionAlgorithms}
    \item database of \textit{patterns}: contains all generated causality patterns, where each pattern has the structure described in Section~\ref{section:concepts:patterns}
\end{itemize}
Furthermore, the cerec system keeps track of all discovered non-causal sentences: as non-causal sentences are not associated with any pattern, they are stored by the system directly, which is of importance for the maintenance of pattern correctness, later described in Section~\ref{section:impl:maintenance}. We use the OpenNLP and Maltparser \cite{nivre2006maltparser} for the NLP unit. 

\subsection{Maintenance of Pattern Correctness}
\label{section:impl:maintenance}

\paragraph{Principles of Maintenance}
In order to ensure the correctness of the database of patterns, a set of principles must be upheld. These principles formalize how the training data -- both causal and non-causal sentences -- constructs the knowledge database represented by the existing patterns. \newline
The most basic principle is the core assumption of the syntactic cause-effect recognition approach: causal sentences of syntactic similarity are likely to have their causal phrases located at the same position within the sentence, which is the foundation of pattern-based detection and extraction algorithms \cite{girju2003automatic}\cite{girju2002text}\cite{fischbach2019automated}. These causal phrases are therefore retrievable by the same phrase extraction methods. This derivation can be denoted as follows:
\begin{multline*}
    \forall s \in S, \exists c \in C, \exists p \in P : causality(s, c) \land compliant(p, s) \rightarrow \\
    applicable(p, s)
\end{multline*}
This principle justifies to have sentences of similar syntactic structure accepted by the same causality pattern. All accepted sentences of a pattern are compliant to the signature of a pattern and therefore share syntactic similarities. This makes it likely for the phrase extraction methods associated with the pattern to extract the correct cause-effect graph. \newline 
The second principle is that every already discovered non-causal sentence must be non-compliant to every pattern, which formally translates to:

$$\forall p \in P, \forall s \in S, \neg \exists c \in C : causality(s,c) \rightarrow \neg compliant(p, s)$$

To ensure that already discovered non-causal sentences are not compliant to a newly introduced pattern, each signature must be specified in a way to comply to both principles of maintenance. The process of achieving this will be technically explained in next section and brought into context in section thereafter. 

\paragraph{Signature Specification}
\label{impl:maintenance:specification}
An important process for the maintenance of pattern correctness in regard to incremental patterns is the specification of a pattern's signature. The incremental nature of the signature, which defines the grammatical class of each pattern and is used to determine a sentence's compliance to it, derives that the structure consists only of as many nodes such that it fulfills all principles of maintenance. \newline 
The overall goal of the machine learning algorithm is to identify all sentences, which are applicable to the same phrase extraction method. To ensure that as many sentences as possible are identified by their respective causality patterns, the pattern's identifying signature must be minimal in size of nodes: each additional node risks that a sentence, which would be applicable by the phrase extraction method of the causality pattern, is not compliant to the pattern anymore, because one node of the sentence's structure differentiates it from the pattern's signature. \newline
The specification process is applied when a sentence is compliant to a pattern, but the sentence is not applicable to the pattern's phrase extraction method, therefore violating the first principle of maintenance. A sentence which is compliant to a pattern but not applicable by its phrase extraction method will be referred to as an \textit{intruder}, because the sentence is falsely compliant to the pattern. An intruder can be causal or non-causal:
\begin{itemize}
    \item Causal intruders are compliant to a pattern but the phrase extraction method of the pattern does not extract the desired cause-effect graph
    \item Non-causal intruders are compliant to a pattern, although they should never be
\end{itemize}
An intruder can only be detected if his actual causality is provided. Knowing the actual causality of a sentence requires semantic domain knowledge and must be provided by a human, either beforehand when preparing training samples or as an correction during the online usage of the system. If an intruder is detected, the pattern's signature must be more specific in order to prevent the pattern's compliance to the intruder. This correction improves the causality detection and extraction of the cerec system in providing further data for the machine-learning algorithm. \newline
The specification of the signature can take two forms: in the most general form, additional nodes are added to the pattern's structure. These nodes are taken from the set of accepted sentences, in order to maintain compliance of the accepted sentences. The algorithm specifically searches for a position in the signature where all accepted sentences contain the same node while the intruding sentence is different. This node is identified as an differentiator between the accepted sentences and the intruder. Adding the differentiator to the pattern's signature will maintain the compliance to the accepted sentences, while revoking the compliance to the intruder. During this process the pattern's signature incrementally reconstructs the structure of the accepted sentences. \newline
The other form of specifying a signature is enforcing additional constraints on certain nodes of the pattern's signature. The implementation explored in our approach uses lexical constraints: explicitly causal cue phrases like "because" or "if" are highly eligible indicators for differentiating between syntactically similar, but causally different sentences. A lexical constraint added to a node of the pattern's signature means that any sentence, that is compliant, must also contain this specific keyword as a word node transitively parented by the node, to which the constraint is applied. \newline
When a signature specification process is triggered, all eligible differentiating specifications are listed and ordered by degree of precision. A specification is more precise when the changes to the signature are minimal, which reduces the risk of later disregarding sentences, which would be accepted but are not compliant anymore due to an overfitting of the signature. The most precise specification is applied.

\paragraph{Training causal Sentences}
\label{impl:maintenance:training_causal}
When training a causal sentence in the cerec system, a sentence alongside a cause-effect graph, which conveys its causality relation, is given. The first step of the system is to check if any existing pattern is compliant to the given sentence. \newline
If no pattern is found, then the given sentence structure is novel to the system and needs to be established. A new pattern is generated in the following steps: first, an identifying signature is associated with the pattern. Following the nature of the incremental signatures, this signature initially consists just of the minimal representation of the structure of its first accepted sentence: the root node of the sentence's structure, which is an S-node identifying the tree as structure of a sentence. \newline
With just this minimal S-node as a signature the new pattern would violate the second principle of maintenance, as all previously discovered non-causal sentences would be falsely compliant to the pattern, since every sentence's structure is rooted in the S-node. This is overcome by applying the specification algorithm described in the previous section: the pattern's signature is specified against every previously discovered non-causal sentence, until no non-causal sentence is compliant to the new pattern and therefore the second principle upheld. \newline
Next, a phrase extraction method is tailored to retrieve the cause-effect graph that conveys the sentence's causality relation from the sentence. This is done by dynamically constructing an algorithm as described in Section~\ref{section:concepts:extractionAlgorithms}. In a last step the given sentence is added as an accepted sentence to the new pattern, as this sentence's affiliation with the pattern is confirmed. \newline
The training algorithm is different when an existing pattern is found to which the causal sentence is compliant. Two cases are possible from here: if the sentence is also applicable to the phrase extraction method associated with the pattern, then this causal sentence is already covered by the cerec system and correctly recognized. The sentence is added to the accepted sentences of the pattern and adds no new pattern to the database. \newline
If the sentence is not applicable to the phrase extraction method associated with the pattern, to which the sentence is compliant, then the sentence violates the first principle of maintenance: the sentence is compliant to a pattern, but not applicable by its phrase extraction methods. \newline
This is resolved by specifying the pattern against the intruding sentence and creating a new pattern for the causal intruder. The result will be two patterns with very similar structures associated to them, but different enough to differentiate the accepted sentences of the original pattern from the intruding sentence. The new pattern will be provided with a specific phrase extraction method to preserve the first principle of maintenance. 

\paragraph{Training non-causal Sentences}
\label{impl:maintenance:training_noncausal}
When training a causal sentence in the cerec system, only a sentence is given. The first step of the system is again to check if any existing pattern is compliant to the given sentence. \newline
If no pattern is found, then the non-causal sentence is correctly discarded. The cerec system adds the sentence to its list of non-causal sentences for further processing. \newline
If a pattern is found, to which the sentence is compliant, then the second principle of maintenance is violated. This is resolved by specifying the structure of the compliant pattern against the non-causal intruder. The specification process results in the incremental addition of nodes or constraints to the pattern's signature, until the intruding sentence is no longer compliant to the pattern and the second principle of maintenance therefore upheld. 

\section{Evaluation}
\label{section:eval}

The purpose of the following preliminary study is to prove the feasibility of the causality recognition and phrase extraction method based on syntactic similarities. For this, we use the cerec system as described in Section~\ref{section:impl} with lexical constraints and train a set of requirements artifacts containing natural language sentences.

\subsection{Design}
\begin{figure*}
    \centering
    \includegraphics[width=0.9\textwidth]{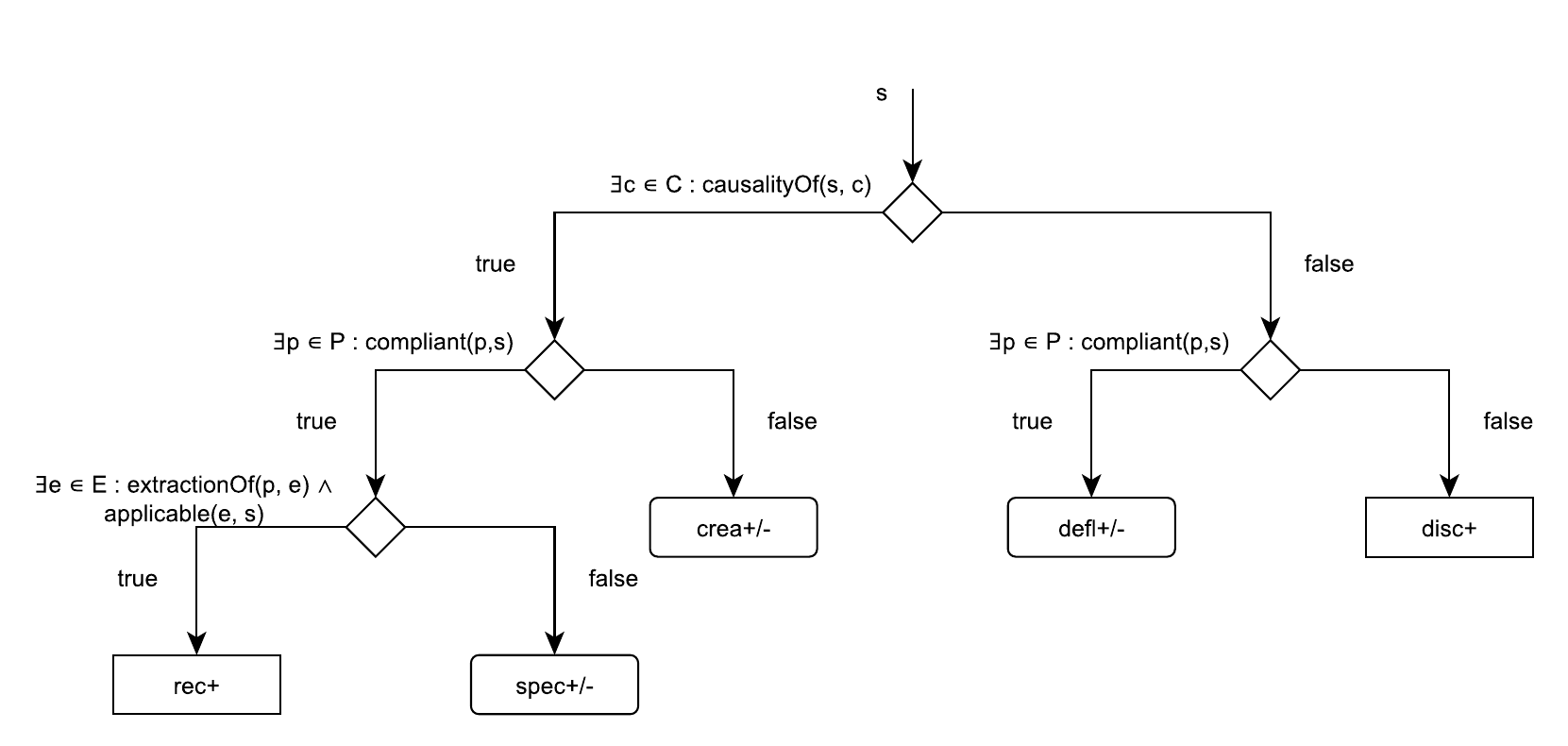}
    \caption{Decision diagram on the training process}
    \label{fig:process:training}
\end{figure*}

A publicly available dataset \textit{A} containing requirements artifacts $a \in A$ was selected for the evaluation of the implemented approach. The obvious choice is the SemEval 2010 Task 8 dataset \cite{hendrickx2010semeval}, which is often used for the evaluation of causality extraction approaches because causal sentences are already annotated. However, the annotated causal phrases are limited to pairs of events, predominantly two nouns, in semantically causal relation. Our proposed algorithm works on the dataset, but evaluating its effectiveness by detecting causally related noun-pairs defeats the purpose of tailoring the proposed algorithm to requirements engineering. Instead we used the PURE dataset of public requirements documents \cite{ferrari2017pure}\cite{pure} and manually annotated the 18 available artifacts for evaluation. Only full natural language sentences were regarded, as these are subject to our approach. The 18 datasets collectively contained 4457 sentences, 558 of which (12.52\%) are causal. \newline
To assess our approach we formulated the following research questions:
\begin{itemize}
    \item \textbf{RQ1}: How effective is the algorithm in automatically detecting and extracting causal relations in a single requirements document?
    \item \textbf{RQ2}: How effective is the algorithm in automatically detecting and extracting causal relations in a single requirements document with previous training?
\end{itemize}
To evaluate the effectiveness of the algorithm, the system will be \textit{fully trained}: a dataset will be randomly ordered and provided to the system sentence by sentence in a \textit{training} operation. The system will perform this operation and evaluate the process with one of the following flags:
\begin{itemize}
    \item creation successful/failed (crea+/crea-): a causal sentence has not been accepted by an existing pattern. The creation of a new pattern is attempted
    %\item creation failed (crea-): a causal sentence has not been accepted by an existing pattern and no new pattern could be established to cover this sentence
    \item recognition successful (rec+): a causal sentence is accepted by an existing pattern and the causal relation has been extracted accordingly
    \item specification successful/failed (spec+/spec-): a sentence has been compliant to an existing pattern, but not applicable to its phrase extraction method, so a pattern specification process to restore the principles of maintenance has been attempted 
    %\item specification failed (spec-): a sentence has been compliant to an existing pattern, but not applicable to its phrase extraction method, and the specification failed to differentiate the sentence from the pattern
    \item discarding successful (disc+): a non-causal sentence has not been accepted by an existing pattern
    \item deflection successful/failed (defl+/defl-): a non-causal sentence has been compliant to an existing pattern, so a pattern specification process to restore the principles of maintenance has been attempted 
\end{itemize}
The eight flags compose the containing set \textit{F}:
$$F = \{rec+, disc+, crea+, crea-, spec+, spec-, delf+, delf-\}$$
The full training process simulates the expected user interaction with the system: while writing natural language sentences of a requirements document, the system attempts to detect a causality in every sentence upon its completion. If a causal sentence is recognized and the causal phrases are extracted correctly, the process is flagged as \textit{rec+}. If the causal sentence is not recognized, the interacting user can manually provide the cause- and effect-phrase of the sentence and train the system by providing the causal relation alongside the sentence. This information is used for a training process and upon success will generate a new pattern, which flags the process as \textit{crea+}. Should this process fail for any reason, the process is flagged as \textit{crea-}. Should the algorithm detect the causality of a sentence but extract a cause- and effect-expression that does not align with the user's perception of the conveyed causality, he can manually correct the cause- and effect-expression of the sentence, which will be used for a specification process as described in Section~\ref{impl:maintenance:specification}, as the sentence is seemingly compliant to a pattern but not applicable by its phrase extraction method. \newline
If the specification succeeds, the run is flagged as \textit{spec+} -- otherwise \textit{spec-}. If a non-causal sentence is non-compliant to any pattern, the sentence is correctly discarded, which is flagged as \textit{disc+}. If the non-causal sentence is wrongfully compliant to a pattern, a specification algorithm has to be performed again to deflect the intruder, but with the exception that no new pattern is created for the new, non-causal sentence. If the specification succeeds, the run is flagged as \textit{defl+} -- otherwise \textit{defl-}. The overall training process is illustrated in Figure~\ref{fig:process:training}. Here processes are written in round-edge boxes and can result in success or failure. \newline 

\subsection{Measures}
The distribution of occurring flags allows to evaluate the effectiveness of the algorithm. We introduce the notation $n_{f}(a)$ as the number of training processes of the artifact $a \in A$ tagged with a specific flag $f \in F$. \newline
The system's task is twofold: the \textit{detection} of a causal relation in a sentence, which is equivalent to a classification as either causal or non-causal, and the \textit{extraction} of a cause-effect graph from a causal sentence. Both tasks are evaluated using the precision, recall and f-score measures. \newline
In case of detection, true positive cases are all processed sentences flagged as \textit{rec+}, \textit{spec+}, and \textit{spec-}, since these are the cases, where a causal sentence is compliant to an existing pattern. The recall value of this task is calculated as described in Equation~\ref{eq:1} and represents, how many of the causal sentences were classified as such.
\begin{equation} \label{eq:1}
    recall(a) = R(a) = \frac{n_{rec+}(a) + n_{spec+}(a) + n_{spec-}(a)}{n_c(a)}
\end{equation}
In case of extraction, true positive cases are all processed sentences flagged as \textit{rec+}. The recall value of this task is calculated as described in Equation~\ref{eq:2} and represents, from how many sentences the cerec system automatically and correctly extracted the cause-effect-graph, hence it is also referred to as \textit{recognition rate}.
\begin{equation} \label{eq:2}
    recall(a) = R(a) = \frac{n_{rec+}(a)}{n_c(a)}
\end{equation}

\subsection{RQ1: Effectiveness of the algorithm without training}
The study of RQ1 simulates the use of the system in an environment without any previous training, which means that no causality patterns exist at the start of the evaluation. Table~\ref{tab:RQ_evaluation} shows the results of the measures on the requirements artifacts on the left side. 

\begin{table*}
    \centering
    \begin{tabular}{|l|r|r|r|r|r|r|r|r|r|r|r|r|r|r|} \hline
        \textbf{Artifact a} & \textbf{n(a)} & \textbf{n\textsubscript{c}(a)} & \multicolumn{6}{c|}{\textbf{RQ1: without previous training}} & \multicolumn{6}{c|}{\textbf{RQ2: with previous training}} \\ \hline
        & & & \multicolumn{3}{c|}{\textbf{Detection}} & \multicolumn{3}{c|}{\textbf{Extraction}} & \multicolumn{3}{c|}{\textbf{Detection}} & \multicolumn{3}{c|}{\textbf{Extraction}} \\ \hline
        & & & \textbf{P} & \textbf{R} & \textbf{F1} & \textbf{P} & \textbf{R} & \textbf{F1} & \textbf{P} & \textbf{R} & \textbf{F1} & \textbf{P} & \textbf{R} & \textbf{F1} \\ \hline
        blitdraft & 67 & 19 & 89.13 & 43.15 & 58.15 & 75.0 & 15.78 & 26.08 & 100.0 & 21.05 & 34.78 & 100.0 & 21.05 & 34.78 \\ \hline
        cctns & 183 & 28 & 87.5 & 20.0 & 32.55 & 84.61 & 15.71 & 26.5 & 100.0 & 14.28 & 25.0 & 100.0 & 7.14 & 13.33 \\ \hline
        dii & 40 & 1 & 0.0 & 0.0 & 0.0 & 0.0 & 0.0 & 0.0 & 0.0 & 0.0 & 0.0 & 0.0 & 0.0 & 0.0 \\ \hline
        eirene & 520 & 66 & 81.57 & 37.57 & 51.45 & 74.31 & 24.54 & 36.9 & 84.73 & 33.63 & 48.15 & 83.47 & 30.6 & 44.78 \\ \hline
        eirene fun & 616 & 140 & 94.0 & 53.71 & 68.36 & 93.4 & 48.57 & 63.9 & 95.81 & 48.99 & 64.83 & 95.44 & 44.85 & 61.03 \\ \hline
        ertms & 223 & 42 & 90.29 & 44.28 & 59.42 & 85.91 & 29.04 & 43.41 & 100.0 & 23.8 & 38.46 & 100.0 & 19.04 & 32.0 \\ \hline
        gammaj & 203 & 3 & 0.0 & 0.0 & 0.0 & 0.0 & 0.0 & 0.0 & 66.66 & 66.66 & 66.66 & 50.0 & 33.33 & 40.0 \\ \hline
        gemini & 392 & 12 & 33.33 & 3.33 & 6.06 & 33.33 & 3.33 & 6.06 & 66.66 & 16.66 & 26.66 & 50.0 & 8.33 & 14.28 \\ \hline
        getreal & 99 & 2 & 0.0 & 0.0 & 0.0 & 0.0 & 0.0 & 0.0 & 0.0 & 0.0 & 0.0 & 0.0 & 0.0 & 0.0 \\ \hline
        keepass & 191 & 46 & 91.3 & 45.65 & 60.86 & 86.66 & 28.26 & 42.62 & 100.0 & 36.95 & 53.96 & 100.0 & 23.91 & 38.59 \\ \hline
        microcare & 39 & 8 & 100.0 & 20.0 & 33.33 & 0.0 & 0.0 & 0.0 & 100.0 & 12.5 & 22.22 & 0.0 & 0.0 & 0.0 \\ \hline
        peering & 131 & 13 & 60.0 & 18.46 & 28.23 & 27.27 & 4.61 & 7.89 & 100.0 & 23.07 & 37.5 & 100.0 & 15.38 & 26.66 \\ \hline
        peppol & 656 & 81 & 71.87 & 17.03 & 27.54 & 56.45 & 8.64 & 14.98 & 95.45 & 10.37 & 18.7 & 87.5 & 3.45 & 6.65 \\ \hline
        phin & 184 & 14 & 72.72 & 11.42 & 19.75 & 40.0 & 2.85 & 5.33 & 86.84 & 47.14 & 61.11 & 54.54 & 8.57 & 14.81 \\ \hline
        qheadache & 107 & 14 & 100.0 & 58.57 & 73.87 & 100.0 & 45.71 & 62.74 & 100.0 & 42.85 & 60.0 & 100.0 & 35.71 & 52.63 \\ \hline
        tcs & 548 & 34 & 65.62 & 12.35 & 20.79 & 45.0 & 5.29 & 9.47 & 67.39 & 18.23 & 28.7 & 55.88 & 11.17 & 18.62 \\ \hline
        themas & 178 & 28 & 97.91 & 67.14 & 79.66 & 97.01 & 46.42 & 62.8 & 100.0 & 31.42 & 47.82 & 100.0 & 21.42 & 35.29 \\ \hline
        video search & 80 & 7 & 91.66 & 31.42 & 46.8 & 83.33 & 14.28 & 24.39 & 100.0 & 14.28 & 25.0 & 100.0 & 14.28 & 25.0 \\ \hline
    \end{tabular}
    \caption{Performance of automatic causality detection and extraction}
    \label{tab:RQ_evaluation}
\end{table*}

\begin{figure}
    \centering
    \includegraphics[width=0.5\textwidth]{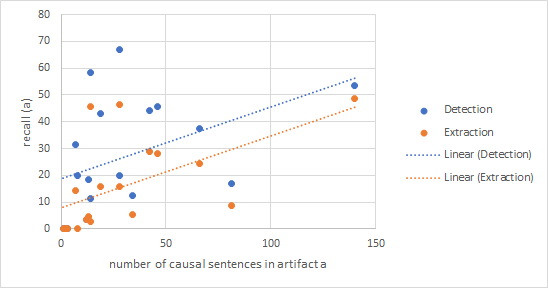}
    \caption{Recall trend in relation to $n_c(a)$}
    \label{fig:recall_trend}
\end{figure}

\begin{figure}
    \centering
    \includegraphics[width=0.5\textwidth]{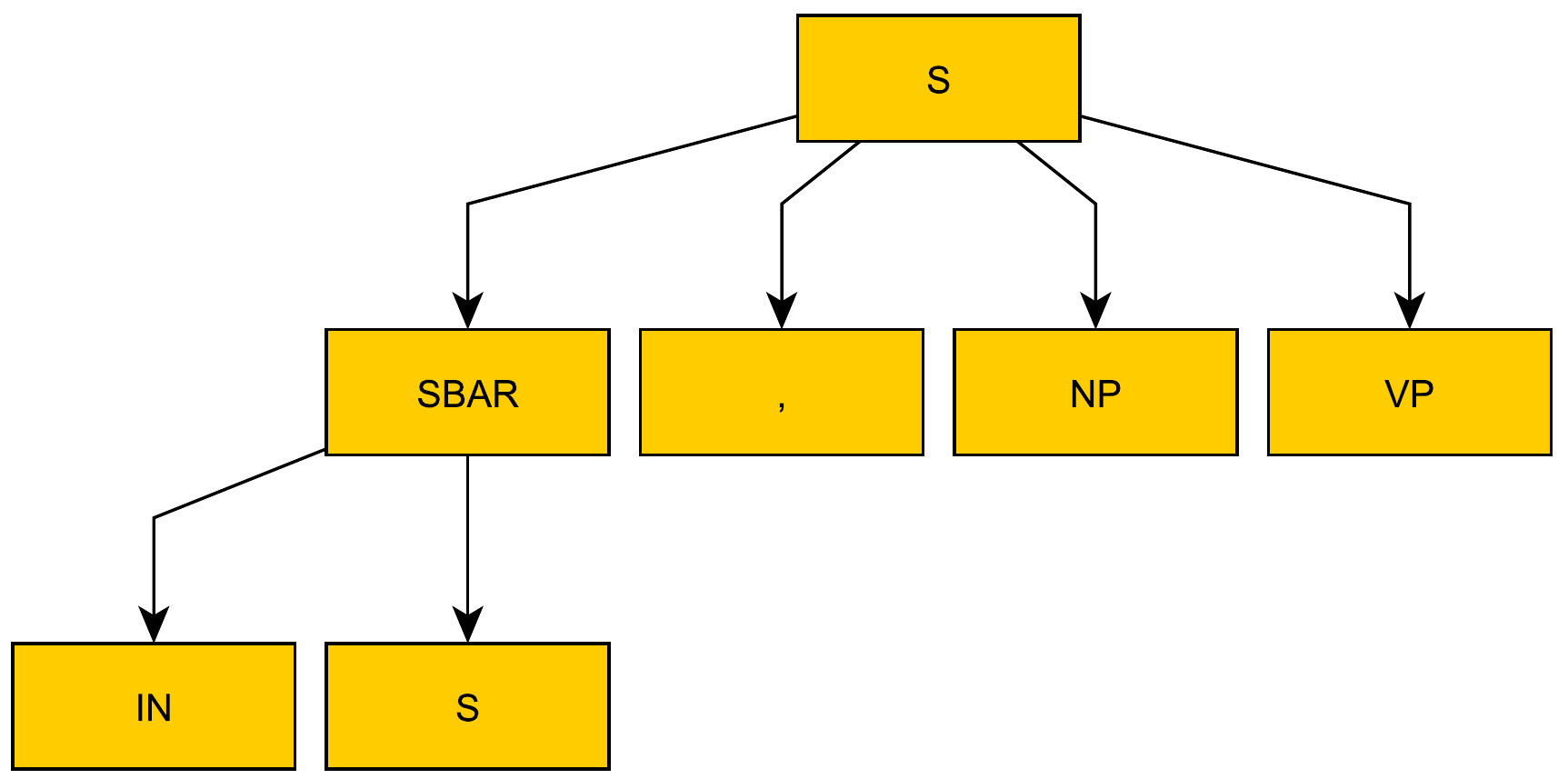}
    \caption{Signature of the most prominent pattern generated from the \textit{qheadache} requirements artifact}
    \label{fig:signature_qheadache}
\end{figure}

Two major insights can be derived from the results of this study: first, the overall number of causal sentences is directly proportionate to the recognition rate. The recall of artifacts with more causal sentences tends to be higher as indicated in Figure~\ref{fig:recall_trend}. The relation between number of causal sentences and recognition rate is due to the nature of the machine learning approach, where an increased number of training data provides more patterns, which the system can learn and reuse for detection and extraction later. \newline
The second insight is that also the number of causal linguistic patterns used is directly proportionate to the recognition rate. This can be observed when comparing the results of the two artifacts \textit{phin} and \textit{qheadache} with each other: even though they contain almost the same amount of causal sentences, qheadache presents a significantly higher precision and recall in both detection and extraction. Investigating the resulting patterns yields that the repeated use of a causal linguistic pattern "If <S>, <NP> <VP>." generated a pattern with the signature shown in Figure~\ref{fig:signature_qheadache}. 
Once the pattern was established by manually annotating the first occurrence of this linguistic pattern, all comparable sentences were compliant and also applicable by the generated phrase extraction method. Sentences like "If the file was correctly updated, there is no output." and "If the game is not saved, a dialog box is displayed that asks to the player if he wants to save the game." are compliant to the pattern and the phrase extraction method generated the correct cause- and effect-phrase, therefore successfully automatizing the causality recognition and extraction for these sentences. \newline
The consequence of the two insights are discussed in the conclusion.

\subsection{RQ2: Effectiveness of the algorithm with previous training}
The study of RQ2 simulates the use of the system in an environment with previous training, which means that causality patterns already exist at the start of the evaluation. To simulate this, each of the requirements artifacts was tested as in RQ1, but only after the system has been previously trained with all other artifacts beforehand, effectively starting the training with all patterns extracted from the other artifacts. The results are shown in Table~\ref{tab:RQ_evaluation} on the right side. \newline
Compared to the results of RQ1, the effect of previous training of the cerec system is not entirely positive: the precision has increased partially, which means that more of the compliant sentences were actually causal and recognized successfully. This comes at the cost of a partially decreased recall, which means that less sentences were compliant to patterns in general. In total these changes result in no overall improvement of the f-score compared to RQ1. \newline
The training beforehand established a large set of patterns based on a corpus vastly outnumbering the sentences contained by the artifact under test. These patterns have undergone a multitude of specification processes as described in Section~\ref{impl:maintenance:specification}, which accounts for the increase in precision of the overall algorithm. On the other hand, the increased specificity of patterns causes less sentences of the artifact under test to be compliant. This represents that the specification process is focused on preventing false positives, hence reducing the recall. \newline
The evaluation of RQ2 shows that the system does not yet profit from an increased beforehand training, which leaves room for improving the scalability of the approach. Reevaluating the focus on precision by balancing the specification process more in favor of recall is to be explored in future approaches. \newline
It is worth mentioning that the lack of improvement in recall when performing RQ2 in contrast to RQ1 is further rooted in the fact that the evaluation was performed on a very heterogeneous set of requirements, written by different authors from different domains. A common template or guideline for writing requirements was most probably absent, which would have increased the similarity of linguistic causality patterns and therefore the recognition rate. The evaluation was from this perspective performed under worst-case conditions with artifacts that are biased against a learning system that relies on consistency and patterned expressions, which mitigates the benefit from previous training.

\section{Conclusion}
\label{section:conclusion}

\subsection{Limitations}
Two major limitations restrict the generalizability of our study approach: first, the approach is restricted to recognizing and extracting a simplified form of cause-effect-graphs, consisting only of a cause- and a effect-phrase, when in reality conveyed causalities may be a lot more complex. Regarding arbitrarily complex cause-effect graphs as formalized causalities improves the applicability of the causality recognition and extraction approach, but may impose a challenge on the cerec system in terms of the recognition rate. \newline
Secondly, the study only investigates the theoretic capability of an automatic causality extraction system while disregarding an actual implementation in practice. Following Roel Wieringa's design science approach \cite{wieringa2014design}, the process framed as \textit{scaling up to practice}, which covers treatment validation, implementation and implementation evaluation, is still required in order to explore the sensitivity of our solution proposal to the practical context. \newline
Further limitations include the exclusive focus on intersentential causality and that patterns' signatures are restricted to lexico-syntactic structures, which disregards the semantic component of causality. The implication of these limitations for the detection and extraction of causal relations from natural language requirements is part of an ongoing study.

\subsection{Discussion}
Though the range of recall values peaks at a promising 48.57\%, some artifacts are evaluated with a recognition rate of 0.0\%. The deviation of success in recognition leads to two observations: first of all, the proposed cerec system lacks in maturity and has still difficulties in maintaining a consistent level of causality recognition. Possible improvements presented in Section~\ref{section:conclusion:future} may improve the robustness of the approach. \newline
Secondly, some requirements artifacts are more eligible for an automatic causality recognition and extraction. Two major factors influence the applicability of a syntactic causality recognition approach:
\begin{enumerate}
    \item Number of causal sentences: a higher number of causal sentences increases the number of causality patterns, which are directly proportionate to the recognition rate
    \item Recurring linguistic causality patterns: reusing formulations for conveying causal relations increases the probability of pattern compliance and phrase extraction method applicability
\end{enumerate}
These derivations are especially interesting when bringing the proposed approach of automatic causality recognition in relation to user interaction: encouraging to comply to the derived recommendations may drastically increase the usage of causal sentences as well as the recognition rate of the cerec system, further benefiting the automatized requirements formalization.

\subsection{Future work}
\label{section:conclusion:future}
Apart from the future work of overcoming the mentioned limitations by utilizing the derivations outlined in the discussion, other aspects of the an automatic causality extraction approach are eligible for further investigation. \newline
The signature of a causality pattern does not only consist of a subtree of a syntactic sentence structure, but may apply additional constraints to increase the specificity of the signature. The possibilities of different constraints and their influence on the recognition rate may yield interesting results on the relation between causal sentences an specific semantic, syntactic and lexical attributes. \newline
Furthermore an integration of the causality recognition and extraction system into a full pipeline dedicated to automatically generating artifacts like test cases from natural language requirements documents may be of interest for validating the context, in which the cerec system is supposed to be used. Research in this direction approaches the application of the causality recognition and extraction and may generate actual value for the requirements engineering phase as well as other downstream phases. \newline
Lastly, the potential of the syntactic patterns might be utilized with different semantic relationships: as the patterns used in our approach simply associate a lexico-syntactic structure to a phrase extraction method retrieving certain parts of the sentence, the algorithm could be tailored towards any semantic relation and context. This may initiate a discussion about the degree of connection between semantic relations and the syntactic structure of sentences. \newline
Overall, the results presented in Section~\ref{section:eval} prove, that the isolated process of automatic causality extraction based on lexico-syntactic patterns holds great potential with an recognition rate of up to 48.57\%, supporting our confidence in the viability of this approach for the requirements engineering context.

%\input{chapters/06_logics}


\begin{acks}
We would like to acknowledge that this work was supported by the KKS foundation through the S.E.R.T. Research Profile project at Blekinge Institute of Technology. We would further like to thank Diego Marmsoler and the reviewers of this paper for their valuable feedback.
\end{acks}

\bibliographystyle{ACM-Reference-Format}
\bibliography{ref.bib}
%\bibliography{sample-base}


\appendix



%Lorem ipsum dolor sit amet, consectetur adipiscing elit. Morbi malesuada, quam in pulvinar varius, metus nunc fermentum urna, id sollicitudin purus odio sit amet enim. Aliquam ullamcorper eu ipsum vel mollis. Curabitur quis dictum nisl. Phasellus vel semper risus, et lacinia dolor. Integer ultricies commodo sem nec semper.

\end{document}